\documentstyle[wider]{article}
\begin{document}
\baselineskip 17pt
\title{{\bf The statistics transmuting Chern-Simons
field and the braid group on Riemann surfaces of genus $g>0$ }}
\author{ Ansar Fayyazuddin \\ \\ \em Institute
of Theoretical Physics \\
\em University of Stockholm \\ \em S-113 46 Stockholm, Sweden}
\date{June 1993  }
\maketitle

\centerline{\bf Abstract}

\bigskip
\noindent
We study bosons interacting with an abelian Chern-Simons field on Riemann
surfaces of genus $g>0$.  It is shown that a singular gauge transformation
brings the hamiltonian to free form.  The transformed wave functions
furnish a multi-component representation of the braid group studied by Imbo and
March-Russell.
The construction constitutes a proof of the equivalence of bosons
coupled to a Chern-Simons field and anyons and generalizes the well known
equivalence of the two pictures on the  plane.
\noindent

USITP-93-16
\hfill
\newpage

\noindent
\section{Introduction}
In 2+1 dimensions particles can have statistics other than Bose or Fermi.
Particles which obey such statistics are called anyons\cite{lm,wil}.  Since the
wave function picks up a complex phase when two particles are exchanged it
is not enough to say that "two particles are exchanged" but one must also
specify {\em how} they are exchanged (clockwise or anti-clockwise,  along a
path enclosing or not enclosing other particles etc.)  This amounts to saying
that the wave function must provide a representation of the braid group.  The
braid group is essentially the group of particle exchanges  which keeps track
of the path of exchange.

It has been known for some time that on the plane one can study the problem of
anyons by coupling bosons to a Chern-Simons field.  The equivalence of the two
pictures is established by constructing a transformation which relates one
picture to the other.  This equivalence has only been
proved for the plane where the global topology of the space is trivial
and for the torus\cite{ans}.
It is the purpose of this paper to establish this equivalence for
space-times of the form $\Sigma_{g}\times R$, where $\Sigma_{g}$ is a
a compact Riemann surface of genus $g>0$.  We know from studies of the
braid group on Riemann surfaces of arbitrary genus that the translation
properties of anyon wave functions become quite complicated as one increases
the genus.  In particular, in addition to the transformations which exchange
particles around homologically trivial paths one also has to take in to
account translations along the handles of the surface $\Sigma_{g}$.
Interestingly, translations around the two homology cycles of a handle do not
in
general commute.  So wave functions can not always provide a single-component
representation of the braid group\cite{einar,imr,wen,lee}.

The paper is structured as follows.  In sect. 2 we collect a few relevant
results from the theory of Riemann surfaces.  In sect. 3 the Chern-Simons
constraint is solved and the remaining degrees of freedom are quantized.
The first quantized hamiltonian is derived in sect. 4 and the dependence
of the wave function on the topological component of the gauge field is
completely determined.  Finally, in sect. 5 we show that the hamiltonian
can be brought to free form by a singular gauge transformation.  The
gauge transformed wave functions furnish a representation of the braid
group studied by Imbo and March-Russell \cite{imr}.

\section{Mathematical Background}
To make the discussion self-contained we state a few mathematical results
here which will be needed in the rest of the paper.  Most of the results
stated here can be found in standard
mathematical references on Riemann surfaces \cite{riem} and in the
physics literature \cite{vv,gaume} where concise and readable accounts
of the main results are presented.
We will work on a compact Riemann surface of genus $g$, which we will denote
by $\Sigma_{g}$ with $g>0$.  We equip $\Sigma_{g}$ with the canonical
homology  basis \{$a_{i}$,$b_{i}$\} shown in Fig. 1.  Associated with this
homology basis is a basis of holomorphic abelian differentials $\omega_{i}
= \mu_{i}\left(z\right)dz$.
They are completely determined by the requirement
\begin{equation}
\int_{a_i}\omega_{j} = \delta_{ij}.
\end{equation}
Their integrals around the $b$ cycles give the "period matrix" $\Omega$
\begin{equation}
\int_{b_i}\omega_{j} = \Omega_{ij}.
\end{equation}
The period matrix $\Omega$ is a $g\times g$, symmetric, complex matrix with
positive  definite imaginary part.  We can form the differentials $\alpha_i$,
$\beta_i$ satisfying
\begin{eqnarray}
{\int_{a_i}}\alpha_{j} & = & \delta_{ij} ,
{\int_{a_i}}\beta_{j} = 0, \;\;\nonumber \\
{\int_{b_i}}\alpha_{j} & = & 0, \;\;
{\int_{b_i}}\beta_{j} = \delta_{ij}.
\end{eqnarray}
by setting
\begin{eqnarray}
\alpha_{i} & = &-\left(\bar{\Omega}\left(\Omega -
\bar{\Omega}\right)^{-1}\right)_{ij}\omega_{j}
 + \left(\Omega\left(\Omega -
\bar{\Omega}\right)^{-1}\right)_{ij}\bar{\omega_{j}} \nonumber \\
\beta_{i} & = &
\left(\Omega - \bar{\Omega}\right)^{-1}_{ij}\left(\omega_{j} -
\bar{\omega_{j}}\right).
\end{eqnarray}
By fixing a point $p_0$ on the surface we can
define the multi-valued integrals
\begin{equation}
\varphi_{i}\left(z\right) = \int_{p_0}^{z}w_{i}.
\end{equation}
These integrals define a mapping $\varphi: \Sigma_{g} \rightarrow
J\left(\Sigma_{g}\right) \simeq C^{g}/L_{\Omega}$ where $L_{\Omega} =
Z^{g} + \Omega Z^{g}$ is the Jacobian lattice associated with the Riemann
surface.  $J\left(\Sigma_{g}\right)$ is called the Jacobian variety of
the Riemann surface $\Sigma_g$.  The mapping $\varphi$ is multi-valued
into $C^g$ but well defined into $J\left(\Sigma_{g}\right)$.  This is so
since two such
mappings differ by an integral $\int \omega$ over a path which is  homologous
to
${\bf m}\cdot{\bf a}+{\bf n}\cdot{\bf b}$ for some ${\bf m},{\bf n} \in Z^{g}$
and therefore the mappings differ by an element of $L_{\Omega}$.
$\varphi\left(\Sigma_{g}\right)$ is a submanifold of
$J\left(\Sigma_{g}\right)$. In the $g=1$ (torus) case this mapping is an
isomorphism.

The surface $\Sigma_g$ comes with a Riemannian metric which in
local coordinates can be written as $ds^{2} = 2\rho\left(z\right) dzd{\bar z}$.

Associated with $\Sigma_g$ and the cohomology basis there is a "prime form"
\cite{vv,mum}
$E\left(z,w\right)$.  The prime form has the important properties of being
holomorphic and odd under $z \leftrightarrow w$.  It is not single-valued
around
the $b$-cycles.  If we transport $z$ around $b_k$ once to $z'$ the prime
form transforms as \cite{vv}:
\begin{equation}
E\left(z',w\right) = -E\left(z,w\right)\exp\left(-i\pi\Omega_{kk} - i2\pi
\int^{w}_{z}\omega_{k}\right). \label{eq:prime}
\end{equation}
The prime form is normalized so that as $z \rightarrow w$, $E\left(z,w\right)
\rightarrow z-w$.  In fact the prime form is the generalization of the
monomials
$z-w$ to Riemann surfaces of higher genus.  The prime form can be written
in terms of the odd theta functions.  The explicit expression will not be
needed here but the interested reader can find it in \cite{vv,mum}.

We will need two real single-valued Green's functions $\log
F\left(z,w\right)$ and $f\left(z,w\right)$ which satisfy
\begin{equation}
\partial_{z}\partial_{\bar{z}}logF\left(z,w\right) =
\pi\delta\left( z-w\right) - \pi\mu_{i}\left(z\right)
\left(Im\Omega\right)^{-1}_{ij}
{\bar \mu}_{j}\left(z\right),
\end{equation}
and
\begin{equation}
\partial_{z}\partial_{\bar{z}}f\left(z,w\right) =
-\pi\delta\left( z-w\right) + \frac{\pi}{A} \rho\left(z\right).
\end{equation}
where $A = \int d^{2}x\rho$ is the area of $\Sigma_g$.  These Green's
functions can be written in terms of the prime form $E\left(z,w\right)$
\cite{vv}
\begin{equation}
F\left(z,w\right) = \exp\left(-i4\pi Im\int_{w}^{z}\omega_{i}
\left(\Omega -\bar{\Omega}\right)^{-1}_{ij} Im\int_{w}^{z}\omega_{j}\right)
\mid E\left(z,w\right)\mid^{2},
\end{equation}
\begin{eqnarray}
f\left(z,w\right) & = & -\log F\left(z,w\right) +\frac{1}{A}\int
d^{2}y \rho \left(\log F\left(z,y\right)+\log F\left(y,w\right)\right)
\nonumber
\\ & - &\frac{1}{A^{2}}\int\int d^{2}xd^{2}y\rho \log F\left(x,y\right).
\end{eqnarray}

\section{Quantization}
The action for non-relativistic bosons coupled to a Chern-Simons field
$a_{\mu}$ is given by:
\begin{equation}
S = \int d^{2}z dt \left(\rho \psi^{\dag}iD_{0}\psi -\frac{1}{2m}
\left({\vec D}\psi\right)^{\dag}\cdot \vec{D}\psi
+\frac{\kappa}{4\pi}\epsilon^{\mu\nu\rho} a_{\mu}\partial_{\nu}a_{\rho}
\right).
\end{equation}
Variation of the action with respect to $a_{0}$ gives the Gauss constraint
\begin{equation}
f_{12} = -\frac{2\pi}{\kappa}J_{0},
\end{equation}
where $J_{0}= \rho\psi^{\dag}\psi$ is the particle number density.
The Gauss law constrains only the curvature associated with the Chern-Simons
connection so it determines the connection upto flat pieces.  I now divide the
gauge potential into three parts: a
flat piece $\theta$, a gauge potential  $a^{\left(1\right)}$ which will be
globally defined on the surface and  therefore has Chern
number $0$, and finally $a^{\left(2\right)}$ which will not be globally well
defined  and will have non-vanishing Chern number \footnote{See also ref.
\cite{salam} where a solution to the Chern-Simons constraint is also given. I
believe, however, that the piece corresponding to my $a^{\left(2\right)}$
is missing.}.  More
precisely $a^{\left(1,2\right)}$ are defined through their curvatures
$f^{\left(1,2\right)}_{12}$:  \begin{equation} f_{12}^{\left(1\right)} =
-\frac{2\pi}{\kappa}\left( J_{0} - \rho \frac{Q}{A}\right),
\end{equation}

\begin{equation}
f_{12}^{\left(2\right)} = -\frac{2\pi Q}{\kappa A}\rho.
\end{equation}
Here $Q= \int d^{2}x J_{0}$ is the particle number operator.
Using the Green's functions defined in the previous section one can
immediately write down the solution for $a^{\left(1\right)}$ (uniquely
upto gauge transformations):
\begin{equation}
a^{\left(1\right)}_{i} = \frac{1}{2\kappa}\epsilon^{ij}\partial_{j}
\int d^{2}w f\left(z,w\right)J_{0}\left(w\right)
\end{equation}
To solve for $a^{\left(2\right)}$, first notice that the volume form
$i\rho dz\wedge d{\bar z}$ is a closed hermitian form therefore
there exists, locally, a function $\phi$ such that $\rho =
\partial{\bar \partial}\phi$ \cite{chern}.  In terms of such a function one can
solve for $a^{\left(2\right)}$:
\begin{equation}
a^{\left(2\right)} = \frac{-i\pi Q}{2\kappa A}\left(\partial \phi dz
- {\bar \partial}\phi d{\bar z}\right)
\end{equation}
Since $a^{\left(2\right)}$ has non-zero Chern number it will not be
globally well defined, but one may pick $\phi$ to be such that
$a^{\left(2\right)}$ is
well defined around the $a$ cycles but not around the $b$ cycles.  Such
a solution is given by
\begin{equation}
\phi = \frac{1}{\pi}\int d^{2}w\log F\left(z,w\right)\rho\left(w\right)
- \frac{A}{2} Im \varphi_{i}\left(z\right)\left(Im \Omega\right)_{ij}
Im \varphi_{j}\left(z\right)
\end{equation}
Finally, we write the flat connection $\theta$ as:
\begin{equation}
\theta = u_{i}\left(t\right)\alpha_{i} + v_{i}\left(t\right)\beta_{i}
\end{equation}
The coefficients $u_{i}, v_{i}$ are the only parts of the connection
not determined by the Gauss law constraint and choice of transition functions
implicitly made in the choice of $a^{\left( 2\right)}$.  They are, therefore,
the only independent degrees of freedom of the gauge field which need to
be quantized.  The part of the action relevant for the quantization of
the $u_{i},v_{i}$ is \footnote{We are assuming that
$\left[u_{i},\psi\right]=\left[v_{i},\psi\right]=0$.}:
\begin{eqnarray}
-\frac{\kappa}{4\pi}\int dt d^{2}z \theta \wedge \dot{\theta}
& = &-\frac{\kappa}{4\pi}\int dt
\left(u_{i}\dot{v}_{j}-\dot{u}_{i}v_{j}\right)
\int_{\Sigma_{g}}\alpha_{i} \wedge \beta_{j} \nonumber \\
& = & -\frac{\kappa}{4\pi}\sum_{k=1}^{g}\int dt
\left(u_{i}\dot{v}_{j}-\dot{u}_{i}v_{j}\right)
\left( \int_{a_k} \alpha_{i} \int_{b_k} \beta_{j}
- \int_{b_k} \alpha_{i} \int_{a_k} \beta_{j} \right) \nonumber \\
& = &\frac{\kappa}{2\pi}\int dt \dot{u}_{i}v_{i} + \mbox{surf. term}
\end{eqnarray}
The second equality is valid for any two closed forms (for a proof see
\cite{riem,gaume}.
Canonical quantization of the $u_{i}$ imposes the commutation relations
\begin{equation}
\left[ u_{i}, v_{j}\right] = i\frac{2\pi}{\kappa}\delta_{ij}
\end{equation}
The usual bosonic commutation relations are valid for the matter fields
$\psi$:
\begin{equation}
\left[\psi,\rho\psi^{\dag}\right]=1.
\end{equation}
All other commutators vanish.

The $u_{i}, v_{i}$ are only defined $\bmod 2\pi$, the rest can be gauged away
by
a single valued gauge transformation.  Even so the wave functions transform
non-trivially under these large gauge transformations whenever $\kappa$ is
not an integer as we shall see in the next section.

\section{The first quantized Hamiltonian and the structure of the Hilbert
Space}
Following ref \cite{jp} we will take the hamiltonian to be
\begin{equation}
{\cal H}= \frac{1}{2m}\left(D_{i}\psi\right)^{\dag}D_{i}\psi
\end{equation}
Let us first define the particle vacuum states annihilated by $\psi$ as:
\begin{equation}
\mid \left\{ u_{i}\right\} > = \mid 0>\otimes_{i=1}^{g}\mid u_{i}>.
\end{equation}
If we write $\kappa = p/q$ with $p,q$ relatively
prime then the {\em physical}\footnote{There is an additional $q^g$ dimensional
degeneracy which is discussed below.}
space of zero particle states is $p^{g}$ dimensional \cite{poly}.

The number operator $Q$ commutes with the hamiltonian, so let us
work in  a particular subspace of the Hilbert space where $Q= N$.  The
Dirac quantization condition requires that $N/\kappa \in Z$, which we
assume from now on.  The
Schr\"{o}dinger wave
function $\Psi$ is defined by
\begin{equation}
\Psi\left({\bf r}_{1},\cdots,{\bf r}_{N};u_{1},\cdots,u_{g}\right)
 =  <\left\{ u_{i}\right\}\mid \psi^{\dag}\left({\bf r}_{1}\right)\cdots
\psi^{\dag}\left({\bf r}_{N}\right)\mid \Psi>
\end{equation}
The Schr\"{o}dinger equation
\begin{equation}
i\frac{\partial}{\partial t}\Psi\left({\bf r}_{1},\cdots,{\bf
r}_{N};u_{1},\cdots,u_{g}\right)
=  <\left\{ u_{i}\right\}\mid\left[\psi^{\dag}\left({\bf
r}_{1}\right)\cdots \psi^{\dag}\left({\bf r}_{N}\right), H\right]\mid \Psi>
\end{equation}
determines the first quantized hamiltonian
\begin{equation}
H^{\left(N\right)} = -\frac{1}{m}\sum_{\alpha = 1}^{N}
\left(D_{\alpha}\bar{D}_{\alpha} + \bar{D}_{\alpha}D_{\alpha}\right),
\end{equation}
where
\begin{eqnarray}
D_{\alpha} & = & \partial_{z_{\alpha}}-ia_{\alpha}, \nonumber \\
\bar{D}_{\alpha} & = & \partial_{\bar{z}_{\alpha}}-i\bar{a}_{\alpha}, \nonumber
\\
a_{\alpha} & = &-u_{i}\left(\bar{\Omega}\left(\Omega -
\bar{\Omega}\right)^{-1}\right)_{ij}\mu_{j}\left(z_{\alpha}\right)
+v_{i}\left(\Omega -
\bar{\Omega}\right)^{-1}_{ij}\mu_{j}\left(z_{\alpha}\right)
\nonumber \\
& &-\frac{i}{2\kappa}\partial_{z_\alpha}\sum_{\beta \neq \alpha} \log
E\left(z_{\alpha}, z_{\beta}\right) + \frac{\pi}{\kappa}\mu_{i}\left(z_{\alpha}
\right)\left(\Omega - \bar{\Omega}\right)^{-1}_{ij}\sum_{\beta = 1}^{N}
\left(\varphi_{j}\left(z_{\beta}\right)-\bar{\varphi}_{j}\left(z_{\beta}\right)
\right).
\end{eqnarray}

Covariance
under large gauge transformations completely determines the dependence of the
wave function on the topological part of the gauge field.  To prove this
we need to make a few definitions.  Let
\begin{eqnarray}
U_{i} & = & \exp \left[
i\kappa v_{i}+i2\pi\sum_{\beta = 1}^{N}\left( \left(\bar{\Omega}
\left(\Omega - \bar{\Omega}\right)^{-1}\right)_{ij}\varphi_{j}\left( z_{\beta}
\right) - \left(\Omega\left(\Omega - \bar{\Omega}\right)^{-1}\right)_{ij}
\bar{\varphi}_{j}\left(z_{\beta}\right)\right)\right], \nonumber \\
V_{i} & = & \exp\left[-i\kappa u_{i} - i2\pi\left(\Omega - \bar{\Omega}
\right)^{-1}_{ij}\sum_{\beta = 1}^{N}\left( \varphi_{j}\left(z_{\beta}
\right) - \bar{\varphi}_{j}\left(z_{\beta}\right)\right)\right].
\end{eqnarray}
$U_{i}, V_{i}, i=1,\cdots,g$ commute with the hamiltonian and generate large
gauge transformations.  They satisfy the commutation relations
\begin{eqnarray}
U_{i}V_{i} & = & V_{i}U_{i}e^{-i2\pi\kappa}, \nonumber \\
 U_{i}V_{j} & = & V_{j}U_{i} \;\;\mbox{$i\neq j$}, \nonumber \\
U_{i}U_{j} & = & U_{j}U_{i}, \nonumber\\
 V_{i}V_{j} & = & V_{j}V_{i}.
\end{eqnarray}
We will require that the wave functions provide an irreducible representation
of
the  group of large gauge transformations.  Each pair $U_{i}, V_{i}$ has a $q$
dimensional irreducible representation.  Let us work in a basis in which
the $V_{i}$ are diagonal
\begin{equation}
V_{i}\Psi_{\bf l} = e^{-i2\pi\kappa l_{i}-i\theta_{1i}}\Psi_{\bf l}.
\end{equation}
The commutation relations then determine the action of the $U_{i}$ upto
a global phase:
\begin{equation}
U_{i}\Psi_{\bf l} = e^{-i\theta_{2i}}\Psi_{l_{1},\cdots,l_{i}-1,\cdots,l_{g}}.
\end{equation}
It is important to realize that the different components of the $q^{g}$
dimensional representation of the group of large gauge transformations
represent the same physical state and therefore the full Hilbert space
contains $q^{g}$ copies of the physical Hilbert space.

In addition to the requirement of providing a representation of the group
of large gauge transformations, the wave function must pick up the correct
gauge transformation whenever a coordinate is translated around a $b$-cycle
of the surface.  This is due to the fact that the gauge field is not
globally defined on $J\left(\Sigma_{g}\right)$.  In particular, translating
$z_\alpha$ around $b_{m}$ once induces:
\begin{equation}
a_{\alpha} \rightarrow a_{\alpha} + \frac{\pi Q}{\kappa}\left(\omega_{m}\left(
z_{\alpha}\right) + \bar{\omega}_{m}\left(z_{\alpha}\right) \right)
\end{equation}
The wave function must compensate this by picking up
a gauge transformation so that
gauge invariant observables are well defined on the surface:
\begin{equation}
\Psi \rightarrow \exp i\frac{\pi Q}{\kappa}\left(\varphi_{m}\left(
z_{\alpha}\right) + \bar{\varphi}_{m}\left(z_{\alpha}\right) \right)\Psi
\end{equation}
The gauge field is periodic under $a$ cycles so the wave function must
also be periodic \footnote{There is nothing preventing the
wave function from picking up an additional global phase under translation
along
either cycle.  Such a phase can easily be included with minor modifications to
the above.  Alternatively, one can set the phase to $0$ and induce a c-number
flat connection in addition to the one above.}

These two requirements along with the mutual commutativity of
$V_{i}, U_{j}^{q}$ and the hamiltonian
determine any wave function to be of the form
\begin{equation}
\Psi_{\bf l} = {\cal G}\sum_{i=1}^{g}\sum_{n_{i}=0}^{p-1}\delta_{{\bf l}
\left\{n_{i}\right\}}F_{\left\{n_{1},\cdots,n_{g}\right\}}\left({\bf r}_{1}
\cdots{\bf r}_{N}\right)
\end{equation}
where
\begin{equation}
\delta_{{\bf l}\left\{n_{i}\right\}} = \prod_{i=1}^{g}
\sum_{j= -\infty}^{\infty}\delta \left(\kappa u_{i} + 2\pi\left(\Omega
-\bar{\Omega}\right)^{-1}_{ik}\sum_{\gamma}\left(\varphi_{k}\left(z_{\gamma}
\right) - \bar{\varphi}_{k}\left(z_{\gamma}\right)\right) -\theta_{1i}
-2\pi\left(j_{i}p+n_{i}\right) - 2\pi\kappa l_{i}\right)
\end{equation}
and
\begin{eqnarray}
{\cal G} & = & \exp -i\left[ \left\{u_{i} + \frac{2\pi}{\kappa}
\sum_{\gamma}\int^{z_{\gamma}}\beta_{i}\right\}
\left\{\frac{\theta_{2i}}{2\pi} -
\sum_{\delta}\int^{z_{\delta}}\alpha_{i}\right\}\right] \nonumber \\
& &\exp \left[\frac{i\pi}{2\kappa}\sum_{\gamma}\int^{z_\gamma}\beta_{j}
\left(\Omega + \bar{\Omega}\right)_{jl}\sum_{\delta}\int^{z_\delta}\beta_{l}
\right]
\end{eqnarray}
$\alpha_{i},\beta_{i}$ are members of the cohomology basis defined in
terms of the abelian differentials $\omega_{i}$ in sect. 2, and the
base point $p_{0}$
should be understood in the integrals .
${\cal G}$ has been factored out for notational convenience.
The $F_{\left\{n_{1},\cdots,n_{g}\right\}}$ are independent of the topological
components $u_i$ of the gauge field and are bosonic functions of the particle
positions.  They satisfy
\begin{equation}
F_{\left\{n_{1},\cdots,n_{g}\right\}}
\left({\bf r}_{1}\cdots{\bf r}_{\alpha}'\cdots
{\bf r}_{N}\right) =
e^{-i\frac{\theta_{1k}}{\kappa}}e^{-i\frac{2\pi}{\kappa}n_{k}}
F_{\left\{n_{1},\cdots,n_{g}\right\}}\left({\bf r}_{1}\cdots{\bf r}_{\alpha}
\cdots{\bf r}_{N}\right), \label{eq:trans1}
\end{equation}
where ${\bf r}_{\alpha}'$ denotes the coordinate ${\bf r}_{\alpha}$ transported
once around the homology cycle $a_{k}$.  Similarly translating
${\bf r}_{\alpha}$ around the cycle $b_{k}$ and denoting it by
${\bf r}_{\alpha}''$
\begin{eqnarray}
& &F_{\left\{n_{1},\cdots,n_{g}\right\}}\left({\bf r}_{1}\cdots{\bf
r}_{\alpha}''\cdots {\bf r}_{N}\right)  =
e^{i\frac{\theta_{2k}}{\kappa}}e^{-i\frac{\pi}{2\kappa}\left(\Omega
+\bar{\Omega}\right)_{kk}}
\exp\left[-i\frac{\pi}{\kappa}\sum_{\gamma}\left(\varphi_{k}\left(z_{\gamma}
\right)+\bar{\varphi}_{k}\left(z_{\gamma}\right)\right)\right] \nonumber \\
& &\exp\left[i\frac{\pi N}{\kappa}\left(\varphi_{k}\left(z_{\alpha}
\right)+\bar{\varphi}_{k}\left(z_{\alpha}\right)\right)\right]
F_{\left\{n_{1}\cdots,n_{k}-1,\cdots n_{g}\right\}}\left({\bf r}_{1}\cdots{\bf
r}_{\alpha}\cdots {\bf r}_{N}\right). \label{eq:trans2}
\end{eqnarray}
The $n_{i}$ are defined $\bmod p$ so $-1 \equiv p-1$ should be understood
in the above expression.

This is the general form of the Schr\"{o}dinger wave function.
No further dependence on the topological components $u_{i}$ is possible
upto terms which can be expressed as gauge transformations.
We have not attempted to solve for the $F_n$ and will need no more than
their transformation properties under translations which are determined
purely by considerations of making gauge invariant observables well
defined on $\Sigma_{g}$.

\section{Gauging away the Chern-Simons interaction}

We now come to the main result of this investigation, which is to prove the
equivalence of anyons and bosons coupled to a Chern-Simons field.  Let us
pause for a moment and discuss what one means by this.  We know that locally
the singular part of the gauge field can be gauged away by a singular gauge
transformation.  This is the well known result of Aharonov and Bohm.  Global
topology plays no role here.  Such a gauge transformation will make
the wave functions multi-valued obeying the correct anyonic statistics under
particle exchange.
However, it
is not enough to gauge away the singular part of the guage field since there
is a residual dependence on the gauge field which corresponds to a non-singular
magnetic field background.  Neglecting the topological components of the
gauge field for a moment, one would conclude that there is no way of
gauging this part of the gauge field away.  Now let us include the topological
components of the gauge field.  How does this change our previous conclusion?
It would appear at first that the above conclusion can not be avoided since
the
topological part of the gauge field is flat and we are still stuck  with a
magnetic field.  But something curious happens when the topological piece is
quantized: we have two different definitions of  the curvature which normally
agree but do not when one has quantized the  flat part of the Chern-Simons
connection.  Specifically, $f_{12} = \partial_{1}a_{2}' -\partial_{2}a_{1}'
\neq 0$ but $F_{12} = \left[D_{1}',D_{2}'\right]=0$! \footnote{The primes refer
to the fact that the singular part of the gauge  field has been removed}.  I
believe that it is this second equality that gives the means to avoid the
depressing conclusion that one would reach otherwise and which is directly
responsible for the possibility of making the  hamiltonian free.

Once we have a free hamiltonian we know that the particles will transform
properly under the appropriate braid group.  We know this to be the case
since Einarsson \cite{einar} and Imbo and March-Russell \cite{imr} have shown
that anyons on  surfaces with genus $g>0$ are multi-component.

Let us see how this happens in practice.  Consider the
gauge transformation
\begin{eqnarray}
U & = & {\cal G}{\cal S} \nonumber \\
{\cal S} & = & \prod_{\gamma <
\delta}\left(\frac{\overline{E\left(z_{\gamma},z_{\delta}
\right)}}{E\left(z_{\gamma},z_{\delta}\right)}\right)^{\frac{1}{2\kappa}}
\end{eqnarray}
${\cal G}$ was defined in the previous section.  ${\cal S}$ is
responsible for removing the singular part of the magnetic field.  The wave
function $\Psi_{\bf l}^{0}$ defined by
\begin{equation}
\Psi_{\bf l} = U\Psi_{\bf l}^{0}
\end{equation}
satisfies the Schr\"{o}dinger equation with respect to the hamiltonian
\begin{eqnarray}
H^{'N} & = & -\frac{1}{m}\sum_{\gamma}\left[\left( \partial_{\gamma}
-\frac{2\pi}
{\kappa}\mu_{i}\left(\Omega - \bar{\Omega}\right)^{-1}_{ij}\frac{\partial}
{\partial u_j}\right)\left(\bar{\partial}_{\gamma} +\frac{2\pi}
{\kappa}\bar{\mu}_{i}\left(\Omega -
\bar{\Omega}\right)^{-1}_{ij}\frac{\partial} {\partial u_j}\right)
\right.\nonumber \\  & + & \left(\bar{\partial}_{\gamma}
+\left.\frac{2\pi} {\kappa}\bar{\mu}_{i}\left(\Omega -
\bar{\Omega}\right)^{-1}_{ij}\frac{\partial} {\partial u_j}\right)\left(
\partial_{\gamma} -\frac{2\pi} {\kappa}\mu_{i}\left(\Omega -
\bar{\Omega}\right)^{-1}_{ij}\frac{\partial} {\partial u_j}\right)\right].
\end{eqnarray}
This hamiltonian is free as can easily be seen by changing variables to
\begin{eqnarray}
\xi_{\gamma} & = & z_{\gamma}, \nonumber \\
\eta_{i} & = & u_{i} + \frac{2\pi}{\kappa}
\left(\Omega - \bar{\Omega}\right)^{-1}_{ij}\sum_{\gamma}\left(
\varphi_{j}\left(z_{\gamma}\right)-\bar\varphi_{j}\left(z_{\gamma}\right)
\right). \label{eq:coord}
\end{eqnarray}
In these variables the hamiltonian is:
\begin{equation}
H^{'N}= -\frac{2}{m}\sum_{\gamma}\frac{\partial}{\partial \xi_{\gamma}}
\frac{\partial}{\partial \bar{\xi}_{\gamma}}.
\end{equation}

Let us first discuss how the transformed wave functions respond to coordinate
exchange.
When we exchange two particle coordinates we must specify whether we
do it in a clockwise or anti-clockwise fashion.  We can keep track of this
by picking the correct analytical continuation of the coordinates on the
Jacobian variety:
$\varphi\left(z_{\alpha}\right)- \varphi\left(z_{\beta}\right)
\rightarrow \exp{\mp i\pi}\left(\varphi\left(z_{\alpha}\right)
-\varphi\left(z_{\beta}\right)\right)$, the minus (plus) sign in the
exponential correspond to clockwise (anti-clockwise) exchange.  Depending
on the choice of the sign in the analytical continuation, the prime form
$E\left( z_{\alpha},z_{\beta}\right)\rightarrow e^{\mp i\pi}
E\left( z_{\beta},z_{\alpha}\right)$.  Denoting clockwise exchange of
particles $\alpha$ and $\beta$ by $\sigma_{\alpha\beta}$ we get
\begin{equation}
\sigma_{\alpha\beta}\Psi_{\bf l}^{0}
= \exp\left(-i\frac{\pi}{\kappa}\right) \Psi_{\bf l}^{0}.
\end{equation}
Here we have used the fact that the $\Psi_{\bf l}$ are bosonic functions of the
particle coordinates.  Notice also that $\sigma_{\alpha\beta}$
acts independently of the particle indices $\alpha , \beta$ so we may write
$\sigma_{\alpha\beta} = \sigma$.

To study the translation properties of the gauge transformed wave function
let us define the following set of wave functions:
\begin{equation}
\Psi_{\bf l}^{m_{1},\cdots,m_{g}} = \prod_{\gamma <
\delta}\left(\frac{\overline{E\left(z_{\gamma},z_{\delta}
\right)}}{E\left(z_{\gamma},z_{\delta}\right)}\right)^{-\frac{1}{2\kappa}}
\sum_{i=1}^{g}\sum_{n_{i}=0}^{p-1}\exp\left(-i\frac{2\pi}{\kappa}n_{i}m_{i}
\right)\delta_{{\bf
l} \left\{n_{i}\right\}}F_{\left\{n_{i}\right\}}\left({\bf r}_{1}\cdots {\bf
r}_{N}\right),
\end{equation}
with $0\leq m_{i} < p$.  For $m_{i} =0$ we recover the wave function
$\Psi_{\bf l}^{0}$ defined above.  In addition we define the translation
operators $A_{k\gamma} \left(B_{k\gamma}\right)$ which translate particle
$\gamma$ once around the $a_{k}$ (resp. $b_k$) cycle of the surface $\Sigma_g$.
{}From eqs (~\ref{eq:prime}),(~\ref{eq:trans1}) and (~\ref{eq:trans2}) one
derives
\begin{equation}
A_{k\gamma}\Psi_{\bf l}^{m_{1},\cdots,m_{g}}
=\exp\left(\frac{-i\theta_{1k}}{\kappa}\right)\Psi_{\bf
l}^{m_{1},\cdots m_{k}+1,\cdots,m_{g}},
\end{equation}
\begin{equation}
B_{k\gamma}\Psi_{\bf l}^{m_{1},\cdots,m_{g}}
=\exp\left(i\frac{\theta_{2k}}{\kappa}
-i\frac{N\pi}{2\kappa}\left(\Omega + \bar{\Omega}\right)_{kk}\right)
\exp\left(-i\frac{2\pi m_{k}}{\kappa}\right)\Psi_{\bf l}^{m_{1},\cdots,m_{g}}.
\end{equation}
It follows that
\begin{eqnarray}
B_{k\beta}A_{k\gamma}& = & A_{k\gamma}B_{k\beta} e^{-i2\pi/\kappa}, \\
B_{l\beta}A_{k\gamma} & = & A_{k\gamma}B_{l\beta} \;\;\mbox{$k\neq l$},
\nonumber \\
A_{k\gamma}A_{l\beta} & = & A_{l\beta}A_{k\gamma} \nonumber \\
B_{k\gamma}B_{l\beta} & = & B_{l\beta}B_{k\gamma} \nonumber \\
\end{eqnarray}
Again $A_{k\gamma}$ and $B_{k\gamma}$ act independently of the particle indices
so we are justified in writing $A_{k\gamma}=A_{k}$ and $B_{k\gamma}=
B_{k}$.  We see that all the relations of the braid group
for spinning particles are satisfied\cite{imr}:
\begin{eqnarray}
\sigma^{2N} & = & 1 \\
B_{k}A_{k} & = & A_{k}B_{k}\sigma^{2},  \nonumber \\
B_{l}A_{k} & = & A_{k}B_{l} \;\;\mbox{$k\neq l$},
\nonumber \\
A_{k}A_{l} & = & A_{l}A_{k}, \nonumber \\
B_{k}B_{l} & = & B_{l}B_{k}, \nonumber \\
\sigma A_{k} & = & A_{k}\sigma, \nonumber \\
\sigma B_{k} & = & B_{k}\sigma. \nonumber \\
\end{eqnarray}
The first of these relations is a consequence of the Dirac quantization
condition.
This proves the equivalence of the braid group description and bosons coupled
to a Chern-Simons field picture.  The way this equivalence comes about is
highly non-trivial as we have seen.  Crucial to the argument was determining
the dependence of the wave function on the topological components of the
gauge field.  This dependence was determined completely by
requiring covariance under
gauge transformations and translation invariance of observables.

It is illuminating to ask what would have happened if
we had not required the wave functions to provide a representation of the
group of large gauge transformations.  The requirement of covariance is really
equivalent to identifying wave functions which are related by large gauge
transformations.  In the absence of such a restriction the Hilbert space
becomes much larger than that described above.  The easiest way to
see this is by looking at the Hilbert space after the gauge transformation
has been performed (the gauge transformation still exists).  The
transformed hamiltonian neither depends on the coordinates $\eta_{i}$
(see eq. (~\ref{eq:coord})) nor on their conjugate momenta.  Therefore,
we can multiply any wave function by arbitrary functions of $\eta_{i}$
and/or superpose wavefunctions with coefficients in the
space of functions of $\eta_{i}$.  We see immediately that such
wave functions have an infinite number of components in general.
When we do require the wave functions to furnish a representation of the
group of large gauge transformations we can not multiply the transformed
wave functions by arbitrary functions of $\eta_i$, since an arbitrary function
will not respect the covariance property.  So we see the crucial role played
by the treatment of $u_{i},v_{i}$ as compact degrees of freedom.

Recently, D. Li \cite{li} has derived the braid group relations and spin of the
anyonic quasi-particle excitations of the fractional quantum hall effect
system.  He has shown that the quasi-particles provide a multi-component
braid group representation similar to the one discussed here.

\section{Acknowledgments}
I thank T. H. Hansson, A. Karlhede, M. Ljungberg, and E. Westerberg for
helpful discussions.  I am grateful to D. Li for communicating his paper
\cite{li} prior to publication.

\end{document}